\documentclass[reqno, eucal]{amsart}

\usepackage[mathscr]{eucal} 
\usepackage[dvips]{graphicx}
\usepackage[dvips]{color}
\usepackage{amsmath,amsfonts,amssymb,amsthm,amscd}
\usepackage{epic}
\usepackage{eepic}
\usepackage{longtable}
\usepackage{array}

\newtheorem{theorem}{Theorem}

\theoremstyle{definition}

\newtheorem{example}[theorem]{Example}
\newtheorem{remark}[theorem]{Remark}

\newcommand{\Z}{{\mathbb Z}}

\newcommand{\C}{{\mathbb C}}

\newcommand{\ok}{{\rm{\bf k}}}

\newcommand{\am}{{\rm{\bf a}}^{\!-} }
\newcommand{\ap}{{\rm{\bf a}}^{\!+} }
\newcommand{\apm}{{\rm{\bf a}}^{\!\pm} }
\newcommand{\amp}{{\rm{\bf a}}^{\!\mp} }

\newcommand{\hf}{{\scriptstyle \frac{1}{2}}}

\begin{document}

\title[Matrix product solutions 
to the reflection equation]{Matrix product solution
to the reflection equation \\
associated with a coideal subalgebra of $\boldsymbol{U_q(A^{(1)}_{n-1})}$}

\author{Atsuo Kuniba}
\address{Atsuo Kuniba, Institute of Physics, 
University of Tokyo, Komaba, Tokyo 153-8902, Japan}
\email{atsuo.s.kuniba@gmail.com}

\author{Masato Okado}
\address{Masato Okado, Department of Mathematics, Osaka City University, 
Osaka, 558-8585, Japan}
\email{okado@sci.osaka-cu.ac.jp}

\author{Akihito Yoneyama}
\address{Akihito Yoneyama, Institute of Physics, 
University of Tokyo, Komaba, Tokyo 153-8902, Japan}
\email{yoneyama@gokutan.c.u-tokyo.ac.jp}

\maketitle

\vspace{0.5cm}
\begin{center}{\bf Abstract}
\end{center}

We present a new solution to the reflection equation
associated with a coideal subalgebra of $U_q(A^{(1)}_{n-1})$
in the symmetric tensor representations and their dual.
Elements of the $K$ matrix are expressed 
by a matrix product formula involving terminating $q$-hypergeometric
series in $q$-boson generators.
At $q=0$, our result reproduces a known set-theoretical solution
to the reflection equation connected to the crystal base theory.

\vspace{0.4cm}

\section{Introduction}
Reflection equation \cite{Ch,Sk,Kul} is a characteristic structure in  
quantum integrable systems in the presence of boundaries.
It combines the $K$ matrix encoding the boundary interaction with 
the $R$ matrix, another   
fundamental object governing the integrability in the bulk \cite{Bax}.
A variety of solutions to the reflection equation have been constructed up to now.
See for example \cite{BFKZ,ML,NR,RV,KP} and references therein. 
In this Letter we present a new solution to the reflection equation
having a number of outstanding features described below.

First, it is associated with the Drinfeld-Jimbo quantum affine algebra 
$U_q(A^{(1)}_{n-1})$ in the symmetric tensor representation $V_{l,z}$ 
and its dual $V^\ast_{l,z}$ with {\em general} degree $l \in \Z_+$.
Here $z$ denotes the (multiplicative) spectral parameter
and $q$ is assumed to be generic throughout.
Both representations $V_{l,z}, V^\ast_{l,z}$ have the 
bases $\{v_\alpha\}$, $\{v^\ast_\alpha\}$
labeled with an array $\alpha=(\alpha_1, \ldots, \alpha_n) \in \Z_+^n$ 
satisfying $\alpha_1+\cdots + \alpha_n=l$.
They include the vector representation as the simplest case $V_{l=1,z}$.
Our $K$ matrix $K(z)=K^{(l)}(z,q)$ is a linear operator 
reflecting the  ``particles" into their duals as
$K(z) : V_{l,z} \rightarrow V^*_{l,z^{-1}}$. 
As such, there are 
three kinds of $R$ matrices 
$R(z), R^\ast(z)$ and $R^{\ast\ast}(z)$ 
(\ref{nam1})--(\ref{nam3}) coming naturally into the game.
They are all well-understood conceptually, 
and admit explicit formulas owing to 
the recent works \cite{KMMO,BM,K}.
The reflection equation takes the form 
\begin{align*}
K_1(x)R^\ast((xy)^{-1})K_1(y)R(xy^{-1})
= 
R^{\ast \ast}(xy^{-1})K_1(y)R^\ast((xy)^{-1})K_1(x),
\end{align*}
where $K_1(x) = K^{(l)}(x,q) \otimes 1$
and 
$K_1(y) = K^{(m)}(y,q) \otimes 1$.
This is an equality of linear maps from 
$V_{l,x} \otimes V_{m,y}$ to 
$V^\ast_{l,x^{-1}} \otimes V^\ast_{m,y^{-1}}$, where
the pair $(l,m) \in \Z_+^2$ is arbitrary.
See (\ref{knh1}) and (\ref{knh2}) for a more concrete description.

Second, let us write the action of our $K$ matrix on the basis as
$K(z) v_\alpha = \sum_\beta K(z)_\alpha^\beta v^\ast_\beta$.
Then, it is 
{\em dense} in the sense that {\em all} the matrix elements 
$K(z)_\alpha^\beta$
are nontrivial rational function of $z$ and $q$.
Put plainly, our $K(z)$ is 
trigonometric, dense, and of type $A$ with general rank $n$ and general ``spin" $l$.
These are distinct features from previous works for type $A$ 
which are mostly devoted to 
diagonal $K$'s or to the situation $\min(n-1,l)=1$\footnote{There are important 
exceptions \cite{KOY,KP} related to this work although.}.

Third, our $K(z)$ is characterized, up to normalization, as the intertwiner of 
the coideal subalgebra $\mathcal{B}_q$ of $U_q(A^{(1)}_{n-1})$ 
generated by the elements
\begin{align*}
b_i = -e_i + q^2 k_i f_i + \frac{q}{1-q}k_i \;\in U_q(A^{(1)}_{n-1}) \qquad (i \in \Z_n).
\end{align*}
Indeed, it is easy to check the right coideal nature
$\Delta\mathcal{B}_q \subset  \mathcal{B}_q\otimes U_q(A^{(1)}_{n-1})$
by applying the coproduct $\Delta$ in (\ref{mhn}) to $b_i$.
The idea to characterize the spectral parameter dependent $K$ matrices in terms of 
coideal subalgebras of quantum affine algebras 
was proposed long ago in the context of 
affine Toda field theory with boundaries.
See for example \cite{DM}, more recent \cite{Ko, RV} and references therein.
Our result may be viewed as a systematic implementation of it for the pair 
$\mathcal{B}_q \subset U_q(A^{(1)}_{n-1})$ 
and the representations $V_{l,z}, V^*_{l,z}$.
We note that the above $b_i$ has also appeared in 
the generalized $q$-Onsager algebra \cite{BB} up to convention.

Last but perhaps most intriguingly, our $K$ matrix has the elements 
that admit an explicit {\em matrix product} formula
\begin{align*}
K(z)^\beta_\alpha = \varrho(z)
\mathrm{Tr}
\bigl(z^{-{\bf h}} \hat{G}^{\beta_1}_{\alpha_1}
\cdots \hat{G}^{\beta_n}_{\alpha_n}\bigr)
\end{align*}
with a scalar $\varrho(z)$.
The trace is taken over a $q$-boson Fock space on which 
${\bf h}$ acts as the number operator.
In terms of the creation $\ap$, the annihilation $\am$ and the $q$-counting generator
$\ok = q^{\bf h}$ of the $q$-boson, the matrix product operator is given as
$\hat{G}^j_i = q^{-\hf i^2}\ok^{-i}G_i^j$ with  
\begin{align*}
G^j_i = (-q;q)_{s}\,
(\ap)^{(j-i)_+}
{}_2\phi_1\Bigl({q^{-t}, -q^{-t} \atop -q^{-s}}; q, 
q \ok \Bigr) (\am)^{(i-j)_+},\quad
s=i+j,\;\; t=\min(i,j),
\end{align*} 
where ${}_2\phi_1$ denotes the $q$-hypergeometric function
and $(m)_+ = \max(m,0)$.
A matrix product solution 
to the reflection equation of this kind was first obtained in \cite{KP}. 
It covered all the fundamental representations of $U_q(A^{(1)}_{n-1})$
whose simplest case goes back to \cite{G}.
According to \cite{KP}, the matrix product structure 
is a signal of three dimensional (3D) integrability.
It is an interesting open problem to elucidate such features 
for the solution in this Letter.
In this regard we note that all the $R$ matrices
appearing in the reflection equation (\ref{sin}) are known to 
admit a matrix product formula 
originating in the tetrahedron equation \cite{K}.

There are further notable properties in our $K$ matrix $K(z)$. 
At $z=q^{-l}$, elements of $K^{(l)}(z,q)$ exhibit a neat factorization (\ref{ask}). 
Combined with the similar property of the $R$ matrices \cite[Th.2]{KMMO},
it allows us to merge the spectral parameter to the spins $l,m \in \Z_+$ 
thereby upgrading the latter to generic parameters.
Consequently we get a parametric generalization of the solution to the reflection equation.
This achieves a boundary analogue of the result 
concerning the Yang-Baxter equation \cite[sec.2.3]{KMMO}.
Another feature of interest occurs at $q=0$, where 
our $K$ matrix and reflection equation (\ref{sin2}) survive quite nontrivially.
In fact they are frozen exactly to 
the set-theoretical (combinatorial) counterparts introduced in 
\cite{KOY} to formulate the box-ball system with reflecting end.

The outline of the Letter is as follows.
In the next section we recapitulate the relevant 
representations of $U_q(A^{(1)}_{n-1})$ and 
the three kinds of $R$ matrices. 
In Section \ref{sec:askb} we introduce the 
coideal subalgebra $\mathcal{B}_q$ and characterize the $K$ matrix
as the intertwiner.
The reflection equation is formulated, which corresponds to a 
twisted one in the terminology of \cite{RV}.
The proof of uniqueness of the intertwiner and the irreducibility of 
$V_{l,x} \otimes V_{m,y}$ as a $\mathcal{B}_q$ module will be given elsewhere.
In Section \ref{sec:ser} we present the matrix product solution to the 
intertwining relation. 
The proof becomes local in the direction of rank, 
and reduces to some quadratic relations of (nonterminating) 
$q$-hypergeometric series.
In Section \ref{sec:ngm} a generalization of integer spins 
(degrees of symmetric tensors and their dual) 
to continuous parameters is described.
In Section \ref{sec:q} we present the results in yet another gauge
and elucidate the connection to the work \cite{KOY} at $q=0$.
Section \ref{sec:d} contains a brief summary and an outlook.
The associated commuting double row transfer matrices (cf. \cite{Sk})
are left for future study.
We set 
$\Z_+ = \Z_{\ge 0}$ and use the following notations:
\begin{align*}
&[u]= \frac{q^u-q^{-u}}{q-q^{-1}},
\;\;(z;q)_m = \prod_{k=1}^m(1-z q^{k-1}),
\;\;\binom{l}{m}_{\!\!q}= \frac{(q;q)_l}{(q;q)_{l-m}(q;q)_m},\\
&\theta(\text{true})=1,\;\;\theta(\text{false}) = 0,\quad
{\bf e}_j = (0,\ldots,0,\overset{j}{1},0,\ldots, 0) \in \Z^n\;\;(1 \le j \le n).
\end{align*}

\section{$U_q(A^{(1)}_{n-1})$ and relevant $R$ matrices}\label{sec:yna}

\subsection{\mathversion{bold}$U_q(A^{(1)}_{n-1})$ and relevant representations}
Let $U_q = U_q(A^{(1)}_{n-1})$ be the Drinfeld-Jimbo quantum affine algebra
(without the derivation operator) 
generated by $e_i, f_i, k^{\pm 1}_i\, (i \in \Z_n)$ obeying the relations
\begin{equation}\label{hn}
\begin{split}
&k_i k^{-1}_i = k^{-1}_i k_i = 1,\;\; [k_i, k_j]=0,\;\;
k_ie_jk^{-1}_i = q^{a_{ij}}e_j,\;\;
k_if_jk^{-1}_i = q^{-a_{ij}}f_j,\;\;
[e_i, f_j]=\delta_{ij}\frac{k_i-k^{-1}_i}{q-q^{-1}},\\
&\sum_{\nu=0}^{1-a_{ij}}(-1)^\nu
e^{(1-a_{ij}-\nu)}_i e_j e_i^{(\nu)}=0,
\quad
\sum_{\nu=0}^{1-a_{ij}}(-1)^\nu
f^{(1-a_{ij}-\nu)}_i f_j f_i^{(\nu)}=0\;\;(i\neq j),
\end{split}
\end{equation}
where $\delta_{ij}= \theta(i=j), \,e^{(\nu)}_i = e^\nu_i/[\nu]!, \,
f^{(\nu)}_i = f^\nu_i/[\nu]!$
and 
$[m]! = \prod_{j=1}^m [j]$.
The Cartan matrix $(a_{ij})_{i,j \in \Z_n}$ is given by 
$a_{ij}= 2\delta_{i,j}-\delta_{i,j+1}-\delta_{i,j-1}$\footnote{Note 
$a_{n-1,0}=a_{0,n-1}=-1$ because of $i,j \in \Z_n$.}.
We employ the coproduct $\Delta$ and the antipode $S$ of the form 
\begin{alignat}{3}
\Delta k^{\pm 1}_i &= k^{\pm 1}_i\otimes k^{\pm 1}_i,&\quad
\Delta e_i &= 1\otimes e_i + e_i \otimes k_i,&\quad
\Delta f_i  &= f_i\otimes 1 + k^{-1}_i\otimes f_i,
\label{mhn}\\
S(k_i) &= k^{-1}_i,  &S(e_i)&= -e_i k^{-1}_i, &
S(f_i) &= -k_i f_i.
\end{alignat}

For integer arrays $\alpha=(\alpha_1,\ldots, \alpha_k),
\beta=(\beta_1,\ldots, \beta_k)  \in \Z^k$ of 
{\em any} length $k$, we use the notation
\begin{align}
|\alpha| &= \sum_{1 \le i \le k}\alpha_i, \quad
\{\alpha \} = \sum_{1 \le i \le k} i \alpha_i,
\quad
\langle \alpha, \beta \rangle = \sum_{1 \le i < j \le k}\alpha_i \beta_j,
\label{air3}\\
\sigma(\alpha) &= 
(\alpha_2,\ldots, \alpha_k,\alpha_1),\qquad
\rho(\alpha) = (\alpha_k, \ldots, \alpha_2, \alpha_1),
\label{aoi}
\end{align}
where $\sigma$ is a cyclic shift and 
$\rho$ is the reverse ordering.
We will be concerned with the two irreducible representations 
of $U_q$ labeled with $l \in \Z_+$:
\begin{align}
\pi_{l,z}: & \;U_q \rightarrow \mathrm{End}(V_{l,z}),
\quad V_{l,z}= \bigoplus_{\alpha \in B_l} \C(q,z)v_\alpha,
\label{obt1}\\
\pi^\ast_{l,z}: & \; U_q \rightarrow \mathrm{End}(V^\ast_{l,z}),
\quad V^\ast_{l,z} = \bigoplus_{\alpha \in B_l} \C(q,z)v^\ast_\alpha,
\label{obt2}
\end{align}
where $B_l$ is a finite set of {\em length $n$} arrays specified as
\begin{align}
B_l = \{\alpha=(\alpha_1, \ldots, \alpha_n) \in \Z_+^n \mid |\alpha| = l\}.
\end{align}
The index $i$ of $\alpha=(\alpha_i)\in B_l$ should always be understood as 
elements of $\Z_n$.
Now the representations (\ref{obt1}) and (\ref{obt2}) are specified as
\begin{alignat}{2}
e_jv_\alpha &= z^{\delta_{j,0}}  [\alpha_{j+1}] 
v_{\alpha+{\bf e}_j-{\bf e}_{j+1}},
& \qquad\quad
e_jv^\ast_\alpha &= -z^{\delta_{j 0}}
[\alpha_{j+1}+1] q^{-\alpha_j+\alpha_{j+1}+2}
v^\ast_{\alpha-{\bf e}_j+{\bf e}_{j+1}},
\label{otm1}\\
f_jv_\alpha &= z^{-\delta_{j 0}} [\alpha_{j}] 
v_{\alpha-{\bf e}_j+{\bf e}_{j+1}},
&
f_jv^\ast_\alpha &= -z^{-\delta_{j 0}} [\alpha_{j}+1]q^{\alpha_j-\alpha_{j+1}}
v^\ast_{\alpha+{\bf e}_j-{\bf e}_{j+1}},
\label{otm2}\\
k_jv_\alpha &= q^{\alpha_j-\alpha_{j+1}}v_\alpha, 
&
k_jv^\ast_\alpha &= q^{-\alpha_j+\alpha_{j+1}}v^\ast_\alpha,
\label{otm3}
\end{alignat}
where $\pi_{l,z}(g), \pi^\ast_{l,z}(g)$ with $g \in U_q$ 
are denoted by $g$ for simplicity. 
In the RHS, $v_\beta, v^\ast_\beta$ with $\beta \not\in B_l$ should be understood as $0$.
The representation $\pi_{l,z}$ is the (affinization of) degree $l$ 
symmetric tensor representation, and 
$\pi^\ast_{l,z}$ is its antipode dual.
Namely,
$(\pi^\ast_{l,z}(g)v^\ast_\alpha, v_\beta) = (v^\ast_\alpha, \pi_{l,z}(S(g))v_\beta)$ holds
for any $\alpha, \beta \in B_l$ and $g \in U_q$ with respect to the bilinear
pairing $(v^\ast_\alpha, v_\beta) = \delta_{\alpha,\beta}$.
In terms of the classical part $U_q(A_{n-1})$, 
they are the irreducible representations  
labeled with the rectangular Young diagrams of shape $1 \times l$ and $(n-1)\times l$,
respectively.

\subsection{\mathversion{bold}$R$ matrices}\label{ss:r}

For simplicity denote the 
tensor product representation
$(\pi^\ast_{l,x} \otimes \pi_{m,y})\circ \Delta$ just 
by $\pi^\ast_{l,x} \otimes \pi_{m,y}$, etc.
Consider the three types of quantum $R$ matrices
which are characterized, up to normalization, by the commutativity with $U_q$ as
\begin{alignat}{2}
R(x/y):& \; V_{l,x} \otimes V_{m,y} \rightarrow V_{m,y} \otimes V_{l,x},
&\qquad 
(\pi_{m,y} \otimes \pi_{l,x})R(x/y)
&= R(x/y)(\pi_{l,x} \otimes \pi_{m,y}),
\label{nam1}\\
R^\ast(x/y):& \; V^\ast_{l,x} \otimes V_{m,y} 
\rightarrow V_{m,y} \otimes V^\ast_{l,x},
&\qquad
(\pi_{m,y} \otimes \pi^\ast_{l,x})R^\ast(x/y)
&= R^\ast(x/y)(\pi^\ast_{l,x} \otimes \pi_{m,y}),
\label{nam2}\\
R^{\ast \ast}(x/y):
& \; V^\ast_{l,x} \otimes V^\ast_{m,y} 
\rightarrow V^\ast_{m,y} \otimes V^\ast_{l,x},
&\qquad 
(\pi^\ast_{m,y} \otimes \pi^\ast_{l,x})R^{\ast\ast}(x/y)
&= R^{\ast\ast}(x/y)(\pi^\ast_{l,x} \otimes \pi^\ast_{m,y}).
\label{nam3}
\end{alignat}
Note that dependence on $l, m, q$ is suppressed in the $R$ matrices.
We specify the matrix elements by 
\begin{align}
R(z)(v_\alpha \otimes v_\beta) 
&= \sum_{\gamma \in B_l, \delta \in B_m}
R(z)^{\gamma, \delta}_{\alpha, \beta}\,v_\delta \otimes v_\gamma,
\\
R^\ast(z)(v^\ast_\alpha \otimes v_\beta) 
&= \sum_{\gamma \in B_l, \delta \in B_m}
R^\ast(z)^{\gamma, \delta}_{\alpha, \beta}\,v_\delta \otimes v^\ast_\gamma,
\\
R^{\ast\ast}(z)(v^\ast_\alpha \otimes v^\ast_\beta) 
&= \sum_{\gamma \in B_l, \delta \in B_m}
R^{\ast\ast}(z)^{\gamma, \delta}_{\alpha, \beta}\,
v^\ast_\delta \otimes v^\ast_\gamma
\end{align}
and the normalization
\begin{align}\label{ymi1}
R(z)^{l{\bf e}_1, m{\bf e}_1}_{l{\bf e}_1, m{\bf e}_1}
= 
R^\ast(z)^{l{\bf e}_1, m{\bf e}_1}_{l{\bf e}_1, m{\bf e}_1}
=
R^{\ast \ast}(z)^{l{\bf e}_1, m{\bf e}_1}_{l{\bf e}_1, m{\bf e}_1}=1.
\end{align}

In order to provide explicit formulas for the $R$ matrices,
we prepare their building blocks.
For complex parameters $\lambda, \mu$ and arrays 
$\beta=(\beta_1, \ldots, \beta_k), 
\gamma=(\gamma_1,\ldots, \gamma_k) \in \Z_+^k$ 
with {\em any} length $k$, define
\begin{align}
\Phi_q(\gamma|\beta; \lambda,\mu) &= 
q^{\langle \beta-\gamma, \gamma\rangle} \left(\frac{\mu}{\lambda}\right)^{|\gamma|}
\overline{\Phi}_q(\gamma|\beta; \lambda,\mu),
\label{rik1}\\
\overline{\Phi}_q(\gamma|\beta; \lambda,\mu) &=
\theta(\gamma \le \beta)
\frac{(\lambda;q)_{|\gamma|}(\frac{\mu}{\lambda};q)_{|\beta|-|\gamma|}}
{(\mu;q)_{|\beta|}}
\prod_{i=1}^{k}\binom{\beta_i}{\gamma_i}_{\!\!q},
\label{rik2}
\end{align}
where $\theta(\gamma \le \beta)$ stands for 
$\prod_{i=1}^k\theta(\gamma_i \le \beta_i)$.
The function $\Phi_q(\gamma|\beta; \lambda,\mu) $ 
was introduced in \cite[eq.(19)]{KMMO}
in the study of a stochastic $R$ matrix for $U_q$.
Following \cite{BM} we 
define a quadratic combination of (\ref{rik1}) as
\begin{align}\label{kan}
A(z)^{\gamma,\delta}_{\alpha,\beta}
&= q^{\langle \alpha, \beta \rangle - \langle \delta, \gamma \rangle}
\sum_{\overline{\xi}+\overline{\eta} = 
\overline{\gamma} + \overline{\delta}}
\Phi_{q^2}(\overline{\xi}-\overline{\delta}|\overline{\xi};
q^{m-l}z, q^{-l-m}z)
\Phi_{q^2}(\overline{\eta}|\overline{\beta};
q^{-l-m}z^{-1},q^{-2m}),
\end{align}
where $\alpha, \gamma \in B_l$ and $\beta, \delta \in B_m$ and 
$\overline{\alpha} = (\alpha_1, \ldots, \alpha_{n-1})$
stands for the truncation of
$\alpha=(\alpha_1,\ldots, \alpha_n)$.
The sum in (\ref{kan}) extends over $\overline{\xi}, \overline{\eta} \in \Z_+^{n-1}$
satisfying $\overline{\xi}+\overline{\eta} = 
\overline{\gamma} + \overline{\delta}$.
There are finitely many such $\overline{\xi}$ and $\overline{\eta}$. 
The function $A(z)^{\gamma,\delta}_{\alpha,\beta}$ satisfies
\begin{align}\label{mkr}
A(z)^{\gamma,\delta}_{\alpha,\beta} = 
A(z)^{\rho(\alpha), \rho(\beta)}_{\rho(\gamma), \rho(\delta)}
\prod_{i=1}^n\frac{(q^2;q^2)_{\alpha_i}(q^2;q^2)_{\beta_i}}
{(q^2;q^2)_{\gamma_i}(q^2;q^2)_{\delta_i}}
= z^{\beta_1-\delta_1}
A(z)^{\sigma(\alpha), \sigma(\beta)}_{\sigma(\gamma), \sigma(\delta)}.
\end{align}
Now the elements of $R$ matrices are expressed as follows
($\delta_\alpha^\beta=\theta(\alpha=\beta)$):
\begin{align}
R(z)^{\gamma, \delta}_{\alpha, \beta} &= 
\delta^{\gamma+\delta}_{\alpha+\beta}A(z)^{\delta, \gamma}_{\beta, \alpha},
\label{rb1}\\
R^\ast(z)^{\gamma, \delta}_{\alpha, \beta} &= 
\delta_{\alpha-\beta}^{\gamma-\delta}
A(z^{-1})^{\rho(\delta), \rho(\alpha)}_{\rho(\beta), \rho(\gamma)},
\label{rb2}\\
R^{\ast\ast}(z)^{\gamma, \delta}_{\alpha, \beta} &= 
\delta_{\alpha+\beta}^{\gamma+\delta}
A(z)_{\rho(\gamma), \rho(\delta)}^{\rho(\alpha), \rho(\beta)}.
\label{rb3}
\end{align}
See the comments after (\ref{LL4}) for the origin of these formulas.
The $R$ matrices satisfy the Yang-Baxter equations \cite{Bax}
reversing the components of the tensor products
$V_{l_1, z_1} \otimes V_{l_2, z_2} \otimes V_{l_3, z_3},
V^\ast_{l_1, z_1} \otimes V_{l_2, z_2} \otimes V_{l_3, z_3},
V^\ast_{l_1, z_1} \otimes V^\ast_{l_2, z_2} \otimes V_{l_3, z_3},
V^\ast_{l_1, z_1} \otimes V^\ast_{l_2, z_2} \otimes V^\ast_{l_3, z_3}$.
In terms of $x=z_1/z_2, y= z_2/z_3$, they read
\begin{align}
(1\otimes R(x))(R(xy) \otimes 1)(1\otimes R(y))
&=
(R(y)\otimes 1)(1\otimes R(xy))(R(x) \otimes 1),
\label{rb4}\\
(1\otimes R^\ast(x))(R^\ast(xy) \otimes 1)(1\otimes R(y))
&=
(R(y)\otimes 1)(1\otimes R^\ast(xy))(R^\ast(x) \otimes 1),
\\
(1\otimes R^{\ast \ast}(x))(R^\ast(xy) \otimes 1)(1\otimes R^\ast(y))
&=
(R^\ast(y)\otimes 1)(1\otimes R^\ast(xy))(R^{\ast\ast}(x) \otimes 1),
\\
(1\otimes R^{\ast \ast}(x))(R^{\ast \ast}(xy) \otimes 1)(1\otimes R^{\ast \ast}(y))
&=
(R^{\ast \ast}(y)\otimes 1)(1\otimes R^{\ast \ast}(xy))(R^{\ast\ast}(x) \otimes 1).
\label{rb5}
\end{align}

\section{A coideal subalgebra and $K$ matrix}\label{sec:askb}

Consider the element
\begin{align}\label{slk}
b_i = -e_i + q^2 k_i f_i + \frac{q}{1-q}k_i \in U_q \qquad (i \in \Z_n)
\end{align}
and let $\mathcal{B}_q$ be the subalgebra of $U_q$ generated by $\{b_i\mid i \in \Z_n\}$.
From  
$\Delta(b_i) = b_i \otimes k_i + 1 \otimes (-e_i + q^2 k_if_i)$, we see 
$\Delta \mathcal{B}_q \subset \mathcal{B}_q \otimes U_q$,
meaning that $\mathcal{B}_q$ is a right coideal subalgebra of $U_q$.
Consider the operator $K(z)=K^{(l)}(z,q)$ 
\begin{align}\label{min}
K(z) \: : \, V_{l,z} \rightarrow V^*_{l,z^{-1}},\qquad
K(z)v_\alpha = \sum_{\gamma \in B_l}
K(z)^\gamma_{\alpha} v^\ast_\gamma,
\end{align}
which satisfies the intertwining relation
\begin{align}\label{fuk}
 K(z)\pi_{l,z}(b) = \pi^\ast_{l, z^{-1}}(b) K(z) \qquad 
(b \in \mathcal{B}_q).
\end{align}
It suffices to impose (\ref{fuk}) for the generators 
$b=b_i\, (i \in \Z_n)$.
From (\ref{otm1})--(\ref{otm3}), it reads explicitly as
\begin{equation}\label{kkn}
\begin{split}
&-z^{\delta_{i0}}[\alpha_{i+1}]K(z)^\gamma_{\alpha+{\bf e}_i-{\bf e}_{i+1}}
+ z^{-\delta_{i0}}[\alpha_i]q^{\alpha_i-\alpha_{i+1}}
K(z)^\gamma_{\alpha-{\bf e}_i+{\bf e}_{i+1}}
+ \frac{1}{1-q}q^{\alpha_i-\alpha_{i+1}+1}K(z)^\gamma_\alpha
\\
&=z^{-\delta_{i0}}q^{-\gamma_i+\gamma_{i+1}}[\gamma_{i+1}]
K(z)^{\gamma+{\bf e}_i-{\bf e}_{i+1}}_\alpha - 
z^{\delta_{i0}}[\gamma_{i}]
K(z)^{\gamma-{\bf e}_i+{\bf e}_{i+1}}_\alpha
+ \frac{1}{1-q}q^{-\gamma_i + \gamma_{i+1}+1}K(z)^\gamma_\alpha,
\end{split}
\end{equation}
where $|\alpha | = |\gamma|=l$ and 
$K(z)^\gamma_{\alpha}=0$ unless $\alpha, \gamma \in B_l$.

The essentials for our construction are the following claim.
\begin{theorem}\label{th:tgm}
The solution $K(z)$ 
to the intertwining relation (\ref{fuk}) or equivalently (\ref{kkn}) 
$(\forall i \in \Z_n)$ is 
unique up to normalization.
Moreover, $V_{l,x} \otimes V_{m,y}$ is irreducible as a 
$\mathcal{B}_q$ module for generic $x$ and $y$.
\end{theorem}

We will prove this for a more general setting elsewhere
based partly on the existence of the crystal base \cite{Kas}.
In what follows, $K(z)$ denotes the unique intertwiner normalized as
\begin{align}\label{knh0}
K(z)^{l{\bf e}_1}_{l{\bf e}_1}=1.
\end{align}

Consider the intertwiner 
$V_{l,x} \otimes V_{m,y} \rightarrow V^\ast_{l, x^{-1}} \otimes V^\ast_{m, y^{-1}}$
of the $\mathcal{B}_q$ modules constructed in two ways as 
\begin{align}
V_{l,x} \otimes V_{m,y}
&\overset{R(xy^{-1})}{\longrightarrow}
V_{m,y} \otimes V_{l,x}
\overset{K_1(y)}{\longrightarrow}
V^\ast_{m, y^{-1}} \otimes V_{l,x} 
\nonumber\\
&\overset{R^\ast((xy)^{-1})}{\longrightarrow}
V_{l,x}\otimes V^\ast_{m, y^{-1}} 
\overset{K_1(x)}{\longrightarrow}
V^\ast_{l, x^{-1}} \otimes V^\ast_{m, y^{-1}},
\label{knh1}\\
V_{l,x} \otimes V_{m,y}
&\overset{K_1(x)}{\longrightarrow}
V^\ast_{l,x^{-1}} \otimes V_{m,y}
\overset{R^\ast((xy)^{-1})}{\longrightarrow}
V_{m,y} \otimes V^\ast_{l, x^{-1}}
\nonumber\\
&\overset{K_1(y)}{\longrightarrow}
V^\ast_{m,y^{-1}} \otimes  V^\ast_{l,x^{-1}}
\overset{R^{\ast\ast}(xy^{-1})}{\longrightarrow}
V^\ast_{l,x^{-1}} \otimes V^\ast_{m,y^{-1}},
\label{knh2}
\end{align}
where $K_1(x) = K^{(l)}(x,q) \otimes 1$ and $K_1(y) = K^{(m)}(y,q) \otimes 1$.
The dependence of each $R$ matrix on $l,m$ should be understood appropriately.
The composition of (\ref{knh1}) and the inverse of (\ref{knh2}) gives a map
on $V_{l,x} \otimes V_{m,y}$ commuting with $\Delta \mathcal{B}_q$.
Then the second assertion in Theorem \ref{th:tgm} 
tells that it must be a scalar multiple of the identity operator.
The scalar is $1$  
due to the normalization (\ref{ymi1}) and (\ref{knh0}).
In this way, we obtain the reflection equation
\begin{align}\label{sin}
K_1(x)R^\ast((xy)^{-1})K_1(y)R(xy^{-1})
= 
R^{\ast \ast}(xy^{-1})K_1(y)R^\ast((xy)^{-1})K_1(x)
\end{align}
of the linear operators 
$V_{l,x} \otimes V_{m,y} \rightarrow 
V^\ast_{l,x^{-1}} \otimes V^\ast_{m,y^{-1}}$
for the intertwiner 
$K(z)$ characterized by the first assertion in Theorem \ref{th:tgm}.
In short, Theorem \ref{th:tgm} achieves {\em linearization}; 
the reflection equation which is 
quadratic in $K(z)$ becomes a corollary of the linear intertwining relation (\ref{fuk}).
In terms of matrix elements (\ref{sin}) reads
\begin{equation}\label{yuk}
\begin{split}
&\sum 
K(x)^{a_3}_{a_2} 
R^\ast((xy)^{-1})^{b_3, a_2}_{b_2, a_1}
K(y)^{b_2}_{b_1}
R(xy^{-1})^{a_1, b_1}_{a_0, b_0}\\
&= \sum 
 R^{\ast\ast}(xy^{-1})^{b_3, a_3}_{b_2, a_2}
K(y)^{b_2}_{b_1}
R^\ast((xy)^{-1})^{a_2, b_1}_{a_1, b_0}
K(x)^{a_1}_{a_0},
\end{split}
\end{equation}
where $a_0, a_3 \in B_l, b_0, b_3 \in B_m$ and the sums range over 
$a_1,a_2 \in B_l, b_1, b_2 \in B_m$ on the both sides.
On the LHS (resp. RHS), they are to obey the weight conservation
$a_1+b_1 = a_0+b_0, a_1-b_2 = a_2-b_3$
(resp. $a_1-b_0 = a_2-b_1, a_2+b_2 = a_3+b_3)$.

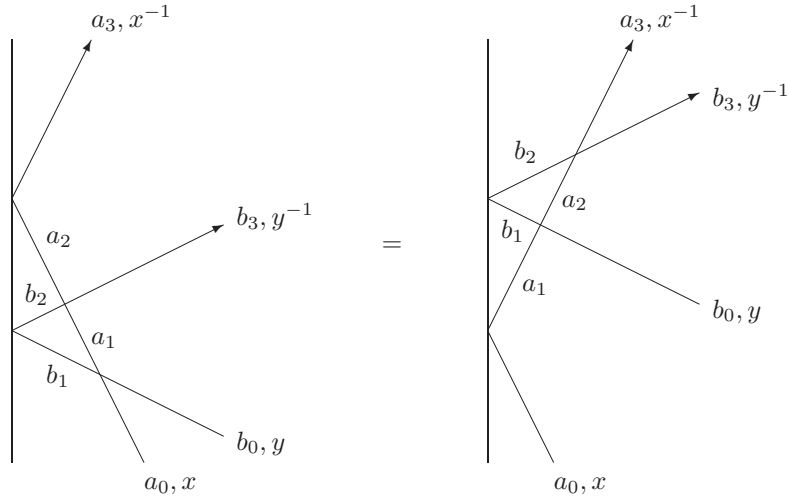
\begin{picture}(300,200)(-70,-20)
\put(0,0){\line(0,1){160}}

\put(0,100){\vector(1,2){30}}\put(30,165){$a_3, x^{-1}$}
\put(0,100){\line(1,-2){50}}\put(50,-10){$a_0, x$}

\put(0,50){\vector(2,1){80}}\put(85,90){$b_3, y^{-1}$}
\put(0,50){\line(2,-1){80}}\put(85,5){$b_0, y$}

\put(13,83){$a_2$}
\put(30,45){$a_1$}

\put(5,61){$b_2$}
\put(13,30){$b_1$}

\put(180,0){
\put(-40,80){$=$}
\put(0,0){\line(0,1){160}}

\put(0,100){\vector(2,1){80}}\put(50,165){$a_3, x^{-1}$}
\put(0,100){\line(2,-1){80}}\put(85,55){$b_0, y$}

\put(0,50){\vector(1,2){55}}\put(85,135){$b_3, y^{-1}$}
\put(0,50){\line(1,-2){25}}\put(25,-10){$a_0, x$}

\put(10,115){$b_2$}
\put(5,85){$b_1$}

\put(13,65){$a_1$}
\put(28,97){$a_2$}

}
\end{picture}

\noindent

\begin{remark}\label{noi}
For the coideal subalgebra generated by  
$-e_i + c_i k_i f_i + d_ik_i$ with $c_id_i \neq 0 \,(\forall i \in \Z_n)$,
a necessary condition for the existence of 
$K(z): \; V_{l,z} \rightarrow V^\ast_{l,w^{-1}}$ with $n\ge 3$ is
\begin{align}
\prod_{i \in \Z_n}c_i = q^{2n}zw^{-1},\qquad d_i^2 = \frac{c_i}{(1-q)^2}.
\end{align} 
Such cases can always be reduced to (\ref{slk}) by applying an algebra
automorphism 
$\omega: e_i \mapsto \mu_i e_i, f_i \mapsto \mu^{-1}_i f_i,
k^{\pm 1}_i \mapsto k_i^{\pm 1}$ of $U_q$ for appropriate constants 
$\mu_i$.
For $n=2$, the intertwiner exists without assuming the right constraint in (39).
However a matrix product formula for such a case is not known in general.
\end{remark}

\section{Matrix product construction}\label{sec:ser}

Let $\mathcal{A}_q$ be the algebra generated by 
$\ap, \am, \ok$ obeying the relations
\begin{align}\label{szk}
\ok \,\ap = q\,\ap \ok,\qquad
\ok \,\am = q^{-1} \am \ok,\qquad
\ap \am = 1- \ok,\qquad
\am \ap = 1- q\ok.
\end{align}
The algebra $\mathcal{A}_q$ will be called $q$-boson.
It is equipped with an anti-algebra automorphism
\begin{align}\label{msk}
\iota: \; \apm \mapsto \amp,\qquad \ok \mapsto \ok.
\end{align}
Let 
$F_q = \bigoplus_{m\ge 0}\C |m\rangle$ 
and $F^\ast_q = \bigoplus_{m \ge 0} \C\langle m |$ be 
the Fock space and its dual 
equipped with the bilinear pairing
$\langle m | m'\rangle = (q; q)_m\delta_{m,m'}$.
They can be endowed with an $\mathcal{A}_q$ module structure by
\begin{equation*}
\begin{split}
&\ap |m\rangle = |m+1\rangle,\quad
\am |m\rangle = (1-q^{m})|m-1\rangle,\quad
\ok |m\rangle = q^{m} |m\rangle,\\
&\langle m | \am = \langle m+1 |, \quad
\langle m | \ap = \langle m-1| (1-q^{m}),\quad
\langle m | \ok = \langle m| q^{m}.
\end{split}
\end{equation*}
It satisfies $(\langle m | X)|m'\rangle = \langle m | (X|m'\rangle)$.
We also use ${\bf h}$ acting on the Fock spaces as 
${\bf h}|m\rangle = m|m\rangle$ and  
$\langle m | {\bf h} = \langle m|m$.
Thus one may set $\ok = q^{\bf h}$.
By definition, the trace on $F_q$ means 
$\mathrm{Tr}(w^{\bf h}X) = 
\sum_{m \ge 0}w^m \frac{\langle m |X|m\rangle}{(q; q)_m}$
when convergent.
The traces appearing in the sequel are always reduced to and evaluated by  
$\mathrm{Tr}(w^{\bf h} \ok^r) = \frac{1}{1-q^rw}$ for
some $w$ and $r \in \Z$ by relation (\ref{szk}).

For each pair $(i,j) \in \Z_+^2$, define an element 
$G^j_i \in \mathcal{A}_q$ by
\begin{equation}\label{aim}
\begin{split}
G^j_i & = (-q;q)_{i+j}\,
\phi\Bigl({q^{-j}, -q^{-j} \atop -q^{-i-j}}; 
q \ok \Bigr)(\am)^{i-j}
\quad (i\ge j),\\
& = (-q;q)_{i+j}\,
(\ap)^{j-i}
\phi\Bigl({q^{-i}, -q^{-i} \atop -q^{-i-j}}; 
q \ok \Bigr)
\quad (i\le j),
\end{split}
\end{equation}
where $\phi$ is a shorthand for the $q$-hypergeometric series
\begin{align}\label{yum}
\phi\Bigl({a, b \atop c}; z \Bigr) = 
{}_2\phi_1\Bigl({a, b \atop c}; q, z \Bigr) = \sum_{m\ge 0}
\frac{(a;q)_m(b;q)_m}{(q;q)_m(c;q)_m}z^m.
\end{align}
The RHS of (\ref{aim}) is terminating and actually involves 
finitely many terms.
Note the properties 
\begin{align}\label{msm}
G^j_i = \iota(G^i_j),\qquad
w^{\bf h}G^j_i = G^j_i\, w^{j-i+{\bf h}}.
\end{align}

\begin{theorem}\label{th:ykn}
The $K$ matrix characterized by (\ref{fuk}) and (\ref{knh0}) 
has the elements expressed by the matrix product formula:
\begin{align}\label{mst}
K(z)^\gamma_\alpha = 
\frac{q^{\langle \gamma, \alpha \rangle}
(q^{-l}z^{-1};q)_{l+1}}{(q^2;q^2)_l (-qz^{-1};q)_l}\,
\mathrm{Tr}\Bigl(
(q^lz)^{-{\bf h}}\,G^{\gamma_1}_{\alpha_1}\cdots G^{\gamma_n}_{\alpha_n}
\Bigr)
\qquad (\alpha, \gamma \in B_l).
\end{align}
\end{theorem}
Due to the right property in (\ref{msm})
and $l^2=|\alpha|^2 = \sum_{i=1}^n \alpha_i^2 + 
2\langle \alpha, \alpha\rangle$ for $\alpha \in B_l$, 
formula (\ref{mst}) is also written as:
\begin{align}\label{mst2}
K(z)^\gamma_\alpha = 
\frac{q^{\frac{1}{2}l^2}
(q^{-l}z^{-1};q)_{l+1}}{(q^2;q^2)_l (-qz^{-1};q)_l}\,
\mathrm{Tr}\Bigl(
z^{-{\bf h}}\,\hat{G}^{\gamma_1}_{\alpha_1}\cdots \hat{G}^{\gamma_n}_{\alpha_n}
\Bigr),
\qquad \hat{G}^j_i = q^{-\frac{1}{2}i^2}\ok^{-i}G^j_i,
\end{align}
where the prefactor of the trace is independent of $\alpha$ and $\gamma$.
Let us sketch a (rather brute force) proof.
Substitute (\ref{mst}) into  (\ref{kkn}).
Applying the right relation in  (\ref{msm}) and  
$
\langle \gamma\pm {\bf e}_i\mp {\bf e}_{i+1}, \alpha \rangle -
\langle \gamma, \alpha \rangle = \pm (\alpha_{i+1}-l\delta_{i0})$,
$\langle \gamma, \alpha\pm {\bf e}_i\mp {\bf e}_{i+1} \rangle - 
\langle \gamma, \alpha \rangle
= \pm (-\gamma_i + l \delta_{i0})$,
we find that (\ref{kkn}) follows from the $\delta_{i0}$-free relation:
\begin{equation}\label{rb}
\begin{split}
&-q^{-\gamma_1}[\alpha_2]
G^{\gamma_1}_{\alpha_1+1}G^{\gamma_2}_{\alpha_2-1}
+q^{\gamma_1+\alpha_1-\alpha_2}[\alpha_1]
G^{\gamma_1}_{\alpha_1-1}G^{\gamma_2}_{\alpha_2+1}
+\frac{q^{\alpha_1-\alpha_2+1}}{1-q}
G^{\gamma_1}_{\alpha_1}G^{\gamma_2}_{\alpha_2}\\
&=
q^{\alpha_2-\gamma_1+\gamma_2}[\gamma_2]
G^{\gamma_1+1}_{\alpha_1}G^{\gamma_2-1}_{\alpha_2}
-
q^{-\alpha_2}[\gamma_1]
G^{\gamma_1-1}_{\alpha_1}G^{\gamma_2+1}_{\alpha_2}
+\frac{q^{-\gamma_1+\gamma_2+1}}{1-q}
G^{\gamma_1}_{\alpha_1}G^{\gamma_2}_{\alpha_2}.
\end{split}
\end{equation}
Substitute (\ref{aim}) into (\ref{rb}) and remove a
common factor after applying the $q$-commutation relations in (\ref{szk}).
Regarding integer powers of $q$ as generic variables,
one is left to show quadratic relations of the 
$q$-hypergeometric series.
Below we illustrate a typical case $\alpha_1 > \gamma_1$ and $\alpha_2 < \gamma_2$.
(The invariance of (\ref{rb}) by $\iota$ in (\ref{msk}) reduces the task in the proof to some extent.)
The relevant quadratic relation reads 
\begin{equation}\label{yne}
\begin{split}
0=&u_1(u_2-u_2^{-1})(-v_1^{-1};q)_2(q^{-1}u_1^2v_1^{-1}w;q)_2\,
\phi\Bigl({u_1, -u_1 \atop -q^{-1}v_1};  w \Bigr)
\phi\Bigl({qu_2, -qu_2 \atop -qv_2}; y \Bigr)\\
&+v_1^{-1}u_2(u_1v_1^{-1}-u_1^{-1}v_1)(-v_2^{-1};q)_2\,
\phi\Bigl({u_1, -u_1 \atop -qv_1};  w \Bigr)
\phi\Bigl({q^{-1}u_2, -q^{-1}u_2 \atop -q^{-1}v_2}; y \Bigr)\\
&-u_1v_2^{-1}(u_2v_2^{-1}-u_2^{-1}v_2)(-v_1^{-1};q)_2\,
\phi\Bigl({q^{-1}u_1, -q^{-1}u_1 \atop -q^{-1}v_1};  w \Bigr)
\phi\Bigl({u_2, -u_2 \atop -qv_2}; y \Bigr)\\
&-u_2(u_1-u^{-1}_1)(-v^{-1}_2;q)_2(q^{-1}u_1^2v_1^{-1}w;q)_2\,
\phi\Bigl({qu_1, -qu_1 \atop -qv_1};  w \Bigr)
\phi\Bigl({u_2, -u_2 \atop -q^{-1}v_2}; y \Bigr)\\
&-(1+q)u_1u_2(v^{-1}_1-v^{-1}_2)(1+v_1^{-1})(1+v_2^{-1})
(1-q^{-1}u_1^2v_1^{-1}w)\,
\phi\Bigl({u_1, -u_1 \atop -v_1};  w \Bigr)
\phi\Bigl({u_2, -u_2 \atop -v_2}; y \Bigr)
\end{split}
\end{equation}
with $y = u_1^2v_1^{-1}u_2^{-2}v_2w$.
Applying Heine's contiguous relations to the factors
$\phi\Bigl({\bullet, \bullet \atop \bullet}; w \Bigr)$,
one can rewrite the RHS as
$A\phi\Bigl({q^{-1}u_1, -u_1 \atop -q^{-1}v_1};  w \Bigr)
+
B\phi\Bigl({u_1, -q^{-1}u_1 \atop -q^{-1}v_1};  w \Bigr)$
with $A,B$ being {\em linear} combinations in 
$\phi\Bigl({\bullet, \bullet \atop \bullet}; y \Bigr)$.
Then it is straightforward, though tedious, to check 
$A=0, B=0$ by (\ref{yum}).
We remark that all the relations like (\ref{yne}) hold
for  generic $u_i, v_i$,  hence for nonterminating $q$-hypergeometric series.

\subsection{Basic properties and examples}
From the matrix product formula (\ref{mst}) it is easy to derive
\begin{align}
K(z)^{\gamma}_{\alpha} &= z^{\alpha_1-\gamma_1}
K(z)^{\sigma(\gamma)}_{\sigma(\alpha)} 
= K(z)^{\rho(\alpha)}_{\rho(\gamma)}, 
\label{mni}\\
K(z)^{\gamma}_{\alpha} &= K(z)^{\gamma^{(i)}}_{\alpha^{(i)}}
\quad \text{if}\;\; \alpha_i = \gamma_i = 0.
\label{sae}
\end{align}
The array $\alpha^{(i)}\in \Z_+^{n-1}$ 
is obtained from $\alpha \in \Z_+^n$ by dropping the $i$th component
$\alpha_i$.
The equality (\ref{sae}) is due to 
$G^0_0= 1$ and 
$\langle \gamma^{(i)}, \alpha^{(i)}\rangle = \langle \gamma, \alpha\rangle$
when $\alpha_i = \gamma_i = 0$.
It implies a reduction with respect to rank $n$
when some components are simultaneously $0$.
In what follows we present the result of an explicit evaluation
of (\ref{mst}) for a few typical cases.
 
\begin{example}\label{mar}
Consider $U_q(A^{(1)}_1)$ for general $l$.
Due to (\ref{mni}), 
$K(z)^{\gamma_1,\gamma_2}_{\alpha_1,\alpha_2}
 = z^{\gamma_2-\alpha_2}
K(z)^{\alpha_1,\alpha_2}_{\gamma_1,\gamma_2}$ holds.
Thus, we present the result assuming 
$s:=\gamma_1-\alpha_1=\alpha_2-\gamma_2 \ge 0$ without loss of 
generality.
\begin{equation}\label{lin}
\begin{split}
K(z)^{\gamma_1,\gamma_2}_{\alpha_1,\alpha_2}
 &=
 q^{\alpha_1\gamma_2}z^{\alpha_1-\gamma_1}\frac{
 (q^{-l}z^{-1};q)_{l+1}(q;q)_s(-q;q)_{\alpha_1+\gamma_1}
 (-q;q)_{\alpha_2+\gamma_2}}{(q^2;q^2)_l(-qz^{-1};q)_l}\\
&\times  \sum_{0 \le j \le \alpha_1}\sum_{0 \le k \le \gamma_2}
\frac{q^{j+k}(q^{-2\alpha_1};q^2)_j(q^{-2\gamma_2};q^2)_k}
{(q^{j+k-l}z^{-1};q)_{s+1}(q;q)_j(q;q)_k 
(-q^{-\alpha_1-\gamma_1};q)_j
(-q^{-\alpha_2-\gamma_2};q)_k}.
\end{split}
\end{equation}
\end{example}

\begin{example}\label{syk}
Consider $U_q(A^{(1)}_{n-1})$  with $l=1$.
The relevant matrix product operators are
\begin{align*}
G^0_0=1, \quad G^1_0 = (1+q)\ap, \quad
G^0_1 = (1+q) \am,\quad
G^1_1 = (1+q)(1+q^2)\Bigl(1-\frac{q(1+q)}{1+q^2}\ok\Bigr).
\end{align*}
Thus, formula (\ref{mst}) yields
\begin{equation}
K(z)^{{\bf e}_j}_{{\bf e}_i} =  
\frac{z^{\delta_{ij}}+q}{z+q} z^{\theta(i<j)}.
\end{equation}
In fact this is the $l=1$ case of more general 
\begin{align}\label{akn} 
K(z)^{l{\bf e}_j}_{l{\bf e}_i} 
= \frac{(-qz^{-\delta_{ij}};q)_l}{(-qz^{-1};q)_l}z^{-l\theta(i>j)}.
\end{align}
\end{example}
 
\begin{example}
Consider $U_q(A^{(1)}_{n-1})$ with $l=2$.
In view of (\ref{mni}) and (\ref{sae}), the matrix elements that are not covered by 
Examples \ref{mar} and \ref{syk} are reduced to the following cases of $n=3$:
\begin{align*}
K(z)^{200}_{011} = \frac{(1+q)^2}{(q+z)(q^2+z)},\;\;
K(z)^{110}_{011} = \frac{(1+q)(1+q+q^2+qz)}{(1+q^2)(q+z)(q^2+z)},\;\;
K(z)^{110}_{101} = \frac{(1+q)(q+z+qz+q^2z)}{(1+q^2)(q+z)(q^2+z)}.
\end{align*}
\end{example}

Let us close the section with the conjecture 
\begin{align}\label{mir}
\lim_{q \rightarrow 0}K(z)^{\gamma}_{\alpha}
&= z^{-Q_0(\gamma,\alpha)},
\end{align}
where 
$Q_0(\gamma,\alpha)$ is defined after (\ref{msa3}).
This indicates that the present gauge 
as well as the one treated in Section \ref{sec:q} also has a curious
connection to the crystal theory \cite{Kas, KMN1, NY, KOY}. 

\section{Parametric generalization}\label{sec:ngm}

\subsection{Factorization at special point}
The function (\ref{rik1}) has two simplifying points:
\begin{align}
\Phi_q(\gamma|\beta; 1,\mu) = \delta_{\gamma,0},\qquad
\Phi_q(\gamma |\beta; \mu,\mu) = \delta_{\gamma,\beta}.
\end{align}
Applying it to (\ref{kan}) and (\ref{rb1})--(\ref{rb3}), we get
\begin{align}
R(q^{m-l})^{\gamma,\delta}_{\alpha,\beta}
&= \delta^{\gamma+\delta}_{\alpha+\beta}\,\theta(\delta\le \alpha)
q^{\langle \beta,\alpha-\delta\rangle + \langle\alpha-\delta,\delta\rangle}
\binom{l}{m}_{\!\!q^2}^{\!\!-1}
\prod_{i=1}^n\binom{\alpha_i}{\delta_i}_{\!\!q^2}\quad (l \ge m),
\label{sar0}
\\
R^\ast(q^{m+l})^{\gamma,\delta}_{\alpha,\beta} 
&= \delta^{\gamma-\delta}_{\alpha-\beta}\,
q^{\langle \delta,\alpha\rangle + \langle \gamma,\beta\rangle}
\binom{l+m}{m}_{\!\!q^2}^{\!\!-1}
\prod_{i=1}^n\binom{\alpha_i+\delta_i}{\alpha_i}_{\!\!q^2},
\\
R^{\ast\ast}(q^{l-m})^{\gamma,\delta}_{\alpha,\beta} 
&= \delta^{\gamma+\delta}_{\alpha+\beta}\,\theta(\alpha\le \delta)
q^{\langle \alpha,\beta-\gamma\rangle + \langle \beta-\gamma,\gamma\rangle}
\binom{m}{l}_{\!\!q^2}^{\!\!-1}
\prod_{i=1}^n\binom{\delta_i}{\alpha_i}_{\!\!q^2}
\quad (l \le m),
\label{sar}
\end{align}
where we assume $\alpha, \gamma \in B_l$ and 
$\beta, \delta \in B_m$ in all the cases.
Up to an overall factor, (\ref{sar}) is due to \cite[Th.2]{KMMO}.
By the argument similar to the proof of it there,
one can show that the $K$ matrix also has the factorization
\begin{align}\label{ask}
K(q^{-l})^\gamma_\alpha = 
\frac{q^{\langle \gamma, \alpha\rangle}
\prod_{i=1}^n(-q;q)_{\alpha_i + \gamma_i}}{(-q;q)_{2l}},
\qquad
K(1)^\gamma_\alpha = 
\frac{\prod_{i=1}^n(-q;q)_{\alpha_i}(-q;q)_{\gamma_i}}
{(-q;q)_l^2}\qquad (\alpha, \gamma \in B_l).
\end{align}

\subsection{\mathversion{bold}Upgrading 
$\lambda = q^{-l}, \mu= q^{-m}$ to generic parameters}

In the reflection equation (\ref{sin}), specialize 
the spectral parameters to $x=q^{-l}, y=q^{-m}$.
Assuming $l\ge m$, one finds that all the $R$ and $K$ matrices 
have the factorized elements given in the previous subsection.
(Note that (\ref{sar}) should be applied after the exchange $l \leftrightarrow m$.) 
Apart from the powers of $q$, (\ref{sar0})--(\ref{sar}) consist of the
$q^2$-multinomial 
$(q^2;q^2)_l/\prod_{i=1}^n(q^2;q^2)_{\alpha_i}
=(-q^{2l-|\overline{\alpha}|+1})^{|\overline{\alpha}|}
(q^{-2l};q^2)_{|\overline{\alpha}|}
/\prod_{i=1}^{n-1}(q^2;q^2)_{\overline{\alpha}_i}$ for $\alpha \in B_l$.
Here $\overline{\alpha}$ is a truncation of $\alpha$ 
explained after (\ref{kan}).
Similar rewriting is possible also for (\ref{ask}).
The powers of $q$ are handled by
$\langle \alpha, \beta \rangle = \langle \overline{\alpha}, \overline{\beta} \rangle
+ |\overline{\alpha}|(m - |\overline{\beta}|)$ for $\beta \in B_m$.
Then from the argument similar to \cite[Sec.2.3]{KMMO},
it follows that the reflection equation, as well as the Yang-Baxter equation,
holds as an identity of a rational function
in which $\lambda = q^{-l}$ and $\mu=q^{-m}$ are regarded as 
generic parameters independent of $q$.
Local spin variables in such a setting range over  
$\overline{\alpha} \in \Z^{n-1}_+$ rather than $\alpha \in B_l$.
Below we describe the resulting $R$ and $K$ matrices resetting 
$\overline{\alpha} \in \Z^{n-1}_+$ to a simpler notation 
$\alpha \in \Z_+^k$.  

For $k \ge 1$, 
introduce the infinite dimensional space
\begin{align} 
W = \bigoplus_{\alpha \in \Z^{k}_+}
\C(q,\lambda, \mu) u_\alpha.
\end{align}
Consider the linear operators 
depending on the continuous parameters $\lambda, \mu$ as
\begin{align}
\mathscr{K}(\lambda) &\in \mathrm{End}(W),\qquad\qquad
\mathscr{R}(\lambda,\mu),\,
\mathscr{R}^\ast(\lambda,\mu),\,
\mathscr{R}^{\ast\ast}(\lambda,\mu) \in \mathrm{End}(W\otimes W),
\\
\mathscr{K}(\lambda) u_\alpha &= \sum_{\gamma \in \Z^{k}_+}
\mathscr{K}(\lambda)_\alpha^\gamma \,u_\gamma,
\qquad 
\mathscr{Q}(\lambda,\mu) (u_\alpha \otimes u_\beta) = 
\sum_{\gamma,\delta \in \Z^{k}_+} 
\mathscr{Q}(\lambda,\mu)_{\alpha,\beta}^{\gamma,\delta}\,
u_\delta \otimes u_\gamma,
\label{skr}
\end{align} 
where $\mathscr{Q} = \mathscr{R},\mathscr{R}^\ast,\mathscr{R}^{\ast\ast}$. 
The matrix elements are defined by
\begin{align}
\mathscr{K}(\lambda)_\alpha^\gamma
&= q^{\langle \gamma,\alpha\rangle + \frac{1}{2}|\alpha|(|\alpha|-1)
+\frac{1}{2}|\gamma|(|\gamma|-1)}
\frac{\prod_{i=1}^{k}(-q;q)_{\alpha_i+\gamma_i}}
{(-\lambda^2;q)_{|\alpha+\gamma|}},
\\
\mathscr{R}^{\ast\ast}(\lambda,\mu)_{\alpha,\beta}^{\gamma,\delta}
&= 
\mathscr{R}
(\mu, \lambda)^{\rho(\beta), \rho(\alpha)}_{\rho(\delta), \rho(\gamma)}
=
\delta_{\alpha+\beta}^{\gamma+\delta}
q^{\langle \beta-\gamma, \gamma\rangle + \langle \alpha, \beta-\gamma\rangle
+|\alpha||\beta|-|\gamma||\delta|}\lambda^{2|\delta-\alpha|}\,
\overline{\Phi}_{q^2}(\alpha|\delta; \lambda^2,\mu^2),
\label{tsk}\\
\mathscr{R}^{\ast}(\lambda,\mu)_{\alpha,\beta}^{\gamma,\delta}
&= \delta_{\alpha-\beta}^{\gamma-\delta}\,
q^{\langle \gamma,\beta\rangle + \langle \delta, \alpha\rangle
+|\alpha||\delta| - |\beta||\gamma|}\,
\overline{\Phi}_{q^2}(\alpha|\alpha+\delta; \lambda^2, \lambda^2\mu^2),
\label{mih}
\end{align}
where $\overline{\Phi}_{q^2}$ is given by (\ref{rik2}).
Then, the Yang-Baxter equations and the reflection equation are valid:
\begin{align}
(1\otimes \mathscr{R}(\lambda,\mu))(\mathscr{R}(\lambda,\nu) \otimes 1)
(1\otimes \mathscr{R}(\mu,\nu))
&=
(\mathscr{R}(\mu,\nu)\otimes 1)(1\otimes \mathscr{R}(\lambda,\nu))
(\mathscr{R}(\lambda,\mu) \otimes 1),
\label{rbb4}
\\
(1\otimes \mathscr{R}^\ast(\lambda,\mu))(\mathscr{R}^\ast(\lambda,\nu) \otimes 1)
(1\otimes \mathscr{R}(\mu,\nu))
&=
(\mathscr{R}(\mu,\nu)\otimes 1)
(1\otimes \mathscr{R}^\ast(\lambda,\nu))
(\mathscr{R}^\ast(\lambda,\mu) \otimes 1),
\\
(1\otimes \mathscr{R}^{\ast\ast}(\lambda,\mu))
(\mathscr{R}^\ast(\lambda,\nu) \otimes 1)
(1\otimes \mathscr{R}^\ast(\mu,\nu))
&=
(\mathscr{R}^\ast(\mu,\nu)\otimes 1)
(1\otimes \mathscr{R}^\ast(\lambda,\nu))
(\mathscr{R}^{\ast\ast}(\lambda,\mu) \otimes 1),
\\
(1\otimes \mathscr{R}^{\ast \ast}(\lambda,\mu))
(\mathscr{R}^{\ast \ast}(\lambda,\nu) \otimes 1)
(1\otimes \mathscr{R}^{\ast \ast}(\mu,\nu))
&=
(\mathscr{R}^{\ast \ast}(\mu,\nu)\otimes 1)
(1\otimes \mathscr{R}^{\ast \ast}(\lambda,\nu))
(\mathscr{R}^{\ast\ast}(\lambda,\mu) \otimes 1),
\label{rbb5}
\\
\mathscr{K}_1(\lambda)\mathscr{R}^\ast(\mu,\lambda)
\mathscr{K}_1(\mu)\mathscr{R}(\lambda, \mu) 
&= 
\mathscr{R}^{\ast\ast}(\mu,\lambda)
\mathscr{K}_1(\mu)
\mathscr{R}^\ast(\lambda,\mu)
\mathscr{K}_1(\lambda).
\end{align}
The Yang-Baxter (resp. reflection)  equations 
hold as identities of the operators on $W^{\otimes 3}$ 
(resp. $W^{\otimes 2}$).
The result (\ref{rbb5}) was obtained 
in \cite[Sec.2.3]{KMMO} up to a gauge of $\mathscr{R}^{\ast\ast}(\lambda, \mu)$. 
Two remarks are in order.

(i) $\mathscr{K}(\lambda)$ and 
$\mathscr{R}^\ast(\lambda, \mu)$ are {\em not} locally finite 
in that the corresponding RHS of (\ref{skr}) contains infinitely many terms.
However, the Yang-Baxter and the reflection equations make sense 
as the identities of matrix elements which are finite for 
any prescribed transitions 
$u_\alpha \otimes u_\beta \otimes u_\gamma \mapsto  
u_{\alpha'} \otimes u_{\beta'} \otimes u_{\gamma'}$
and 
$u_\alpha \otimes u_\beta \mapsto  
u_{\alpha'} \otimes u_{\beta'}$.

(ii) The Yang-Baxter equations (\ref{rbb4}) -- (\ref{rbb5}) remain valid 
under the replacement 
\begin{align}
\mathscr{R}(\lambda,\mu)_{\alpha, \beta}^{\gamma, \delta} 
&\mapsto 
q^{\varphi_1(\delta,\gamma) - \varphi_1(\alpha,\beta)}
\bigl(\lambda/\mu\bigr)^{\varphi_2(\gamma-\alpha)}
\mathscr{R}(\lambda,\mu)_{\alpha, \beta}^{\gamma, \delta},
\label{air1}\\
\mathscr{R}^\ast(\lambda,\mu)_{\alpha, \beta}^{\gamma, \delta} 
&\mapsto 
q^{\varphi_1(\alpha, \delta) - \varphi_1(\beta,\gamma)}
\lambda^{\varphi_3(\gamma-\alpha)}\mu^{\varphi_2(\delta-\beta)}
\mathscr{R}^\ast(\lambda,\mu)_{\alpha, \beta}^{\gamma, \delta},
\\
\mathscr{R}^{\ast\ast}(\lambda,\mu)_{\alpha, \beta}^{\gamma, \delta} 
&\mapsto 
q^{\varphi_1(\beta, \alpha) - \varphi_1(\gamma,\delta)}
\bigl(\lambda/\mu\bigr)^{\varphi_3(\gamma-\alpha)}
\mathscr{R}^{\ast\ast}(\lambda,\mu)_{\alpha, \beta}^{\gamma, \delta},
\label{air2}
\end{align}
where $\varphi_1$ (resp. $\varphi_2, \varphi_3$) is 
any bilinear (resp. linear) function. 
This can be utilized to simplify 
(\ref{tsk}) and (\ref{mih}) to some extent.
However there is no bilinear function 
$\varphi'(\cdot, \cdot)$ such that the 
transformation 
$\mathscr{K}(\lambda)_\alpha^\gamma
\mapsto q^{\varphi'(\gamma, \alpha)}\mathscr{K}(\lambda)_\alpha^\gamma$
combined with (\ref{air1})--(\ref{air2}) 
preserves the reflection equation.

\section{Another gauge}\label{sec:q}

The results in Sections \ref{sec:yna} and \ref{sec:askb} can also be stated in another gauge 
which suits the study of the limit $q \rightarrow 0$
in relation to the crystal theory \cite{Kas}.

\subsection{\mathversion{bold}Representation 
$\pi^\vee_{l,z}$ and associated $R$ matrices}
Consider the representation (\cite[eq.(2)]{KMMO}, \cite[eq.(3.14)]{K})
\begin{align}
\pi^\vee_{l,z}: & \; U_q \rightarrow \mathrm{End}(V^\vee_{l,z}),
\qquad V^\vee_{l,z} = \bigoplus_{\alpha \in B_l} \C(q,z)v^\vee_\alpha,
\label{kyk}\\
e_jv^\vee_\alpha &= z^{\delta_{j 0}}
[\alpha_j] v^\vee_{\alpha-{\bf e}_j+{\bf e}_{j+1}},
\quad
f_jv^\vee_\alpha = z^{-\delta_{j 0}} [\alpha_{j+1}]
v^\vee_{\alpha+{\bf e}_j-{\bf e}_{j+1}},
\quad
k_jv^\vee_\alpha = q^{-\alpha_j+\alpha_{j+1}}v^\vee_\alpha,
\end{align}
where again $\pi^\vee_{l,z}(g)$ are abbreviated to $g$.
It is easy to see the equivalence 
\begin{align}
\pi^\ast_{l,z} \simeq \pi^\vee_{l,(-q)^n z}\quad
\text{via the identification}\;\;
v^\vee_\alpha = (-q)^{\{\alpha\}}\prod_{i=1}^n(q^2;q^2)_{\alpha_i}v^\ast_\alpha
\end{align}
by means of $\{\alpha\pm ({\bf e}_i - {\bf e}_{i+1})\}
-\{\alpha\} = \pm(-1+n\delta_{i0})$.
See (\ref{air3}) for the definition of the symbol $\{\alpha\}$.
Denote 
the counterparts of the $R$ matrices in (\ref{nam2}) and (\ref{nam3})  by 
\begin{alignat}{2}
R^\vee(x/y):& \; V^\vee_{l,x} \otimes V_{m,y} 
\rightarrow V_{m,y} \otimes V^\vee_{l,x},
&\qquad
(\pi_{m,y} \otimes \pi^\vee_{l,x})R^\vee(x/y)
&= R^\vee(x/y)(\pi^\vee_{l,x} \otimes \pi_{m,y}),
\label{LL2}\\
R^{\vee \vee}(x/y):
& \; V^\vee_{l,x} \otimes V^\vee_{m,y} 
\rightarrow V^\vee_{m,y} \otimes V^\vee_{l,x},
&\qquad 
(\pi^\vee_{m,y} \otimes \pi^\vee_{l,x})R^{\vee\vee}(x/y)
&= R^{\vee\vee}(x/y)(\pi^\vee_{l,x} \otimes \pi^\vee_{m,y}).
\label{LL3}
\end{alignat}
Under the normalization
$R^\vee(z)^{l{\bf e}_1, m{\bf e}_1}_{l{\bf e}_1, m{\bf e}_1}
=
R^{\vee \vee}(z)^{l{\bf e}_1, m{\bf e}_1}_{l{\bf e}_1, m{\bf e}_1}=1$
as in (\ref{ymi1}),
their matrix elements are given by 
\begin{align}\label{LL4}
R^\vee(z)_{\alpha, \beta}^{\gamma,\delta} = 
\delta^{\gamma-\delta}_{\alpha-\beta}
(-q)^{\{\beta-\delta\}}\prod_{i=1}^n
\frac{(q^2;q^2)_{\beta_i}}{(q^2;q^2)_{\delta_i}}
A((-q)^nz^{-1})_{\delta, \alpha}^{\beta, \gamma},
\qquad
R^{\vee\vee}(z)_{\alpha, \beta}^{\gamma,\delta} = 
\delta^{\gamma+\delta}_{\alpha+\beta}
A(z)_{\alpha, \beta}^{\gamma,\delta}.
\end{align}
The above formula for $R^{\vee\vee}(z)_{\alpha, \beta}^{\gamma,\delta}$ 
was obtained in \cite{BM} extending the 
result of \cite{KMMO}.
The one for $R^\vee(z)_{\alpha, \beta}^{\gamma,\delta}$ and (\ref{rb1})--(\ref{rb3}) 
can be deduced from it by applying the crossing symmetry and 
the results in \cite{K} especially eqs.(2.7), (2.42) and  Th.3.1 therein.
The Yang-Baxter equations (\ref{rb4})--(\ref{rb5}) with $\ast$ replaced by $\vee$
are valid.

\subsection{\mathversion{bold}$K$ matrix and reflection equation}
From now on, we set 
\begin{align*}
q=-p^2
\end{align*}
but allow coexistence of $q$ and $p$ when it eases the presentation.
Let $\mathcal{B}'_q$ be the right coideal subalgebra of $U_q$ generated by 
\begin{align}\label{syr}
b'_i = e_i + q k_i f_i + \frac{p}{1-q}k_i \in U_q\qquad (i \in \Z_n).
\end{align}
This is related to $b_i$ in (\ref{slk}) via
$b'_i = -p^{-1}\omega(b_i)$ where $\omega$ denotes the automorphism
mentioned in Remark \ref{noi} with $\forall \mu_i = p$.
Let 
\begin{align}\label{nzm1}
K'(z) \; : \, V_{l,z} \rightarrow V^\vee_{l,z^{-1}},\qquad
K'(z)v_\alpha = \sum_{\gamma \in B_l}
K'(z)^\gamma_{\alpha} v^\vee_\gamma
\end{align}
be the unique map satisfying the intertwining relation 
\begin{align}\label{nzm2}
\quad K'(b)\pi_{l,z}(b) = \pi^\vee_{l, z^{-1}}(b) K'(z) \qquad
(b \in \mathcal{B}'_q)
\end{align}
and the normalization 
$K'(z)^{l{\bf e}_1}_{l{\bf e}_1}=1$.
From the construction so far we find that its matrix elements are related to 
those of $K(z)$ as 
\begin{align}\label{ykw}
K'(z)^\gamma_\alpha = p^{\{\alpha-\gamma\}}
\frac{(q^2;q^2)_l}{\prod_{i=1}^n(q^2;q^2)_{\gamma_i}}
K(p^n z)^\gamma_\alpha\qquad(\alpha, \gamma \in B_l).
\end{align}
Similarly to (\ref{sin}), it satisfies the reflection equation
\begin{align}\label{sin2}
K'_1(x)R^\vee((xy)^{-1})K'_1(y)R(xy^{-1})
= 
R^{\vee \vee}(xy^{-1})K'_1(y)R^\vee((xy)^{-1})K'_1(x)
\end{align}
as linear operators 
$V_{l,x} \otimes V_{m,y} \rightarrow 
V^\vee_{l,x^{-1}} \otimes V^\vee_{m,y^{-1}}$.

\subsection{\mathversion{bold}Combinatorial $R$ and $K$ at $q=0$}

At $q=0$, the $R$ matrices survive nontrivially as
\begin{align}
\lim_{q\rightarrow 0}R(z)^{\gamma, \delta}_{\alpha,\beta}
&= \theta\bigl(\mathrm{R}(\beta \otimes \alpha) = \gamma \otimes \delta\bigr)\,
z^{-Q_0(\beta, \alpha)}, 
\label{msa1}\\
\lim_{q\rightarrow 0}R^\vee(z)^{\gamma, \delta}_{\alpha,\beta}/
R^\vee(z)^{l{\bf e}_1, m{\bf e}_2}_{l{\bf e}_1, m{\bf e}_2}
&= \theta\bigl(\mathrm{R}^{\vee}(\beta \otimes \alpha) 
= \gamma \otimes \delta\bigr)\,
z^{-P_{0}(\beta, \alpha)}, 
\label{msa2}\\
\lim_{q\rightarrow 0}R^{\vee\vee}(z)^{\gamma, \delta}_{\alpha,\beta}
&= \theta\bigl(\mathrm{R}^{\vee\vee}
(\beta \otimes \alpha) = \gamma \otimes \delta \bigr)\,
z^{-Q_0(\alpha, \beta)}, 
\label{msa3}
\end{align}
where
$P_i(\alpha, \beta) = \min(\alpha_{i+1},\beta_{i+1})$, 
 $Q_i(\alpha, \beta) = \min_{1 \le k \le n} 
\Bigl\{\sum_{1 \le j < k}\alpha_{i+j} + \sum_{k< j \le n}\beta_{i+j}\Bigr\}$.
The denominator in the second formula is given by 
$R^\vee(z)^{l{\bf e}_1, m{\bf e}_2}_{l{\bf e}_1, m{\bf e}_2}
=((-q)^{1-n}z)^m \frac{(q^{l-m+n}z^{-1};q^2)_m}
{(q^{l-m-n+2}z;q^2)_m}$ from (\ref{LL4}).
In the RHS, 
we regard $\alpha, \gamma \in B_l, 
\beta, \delta \in B_m$ 
as elements of {\em crystals} \cite{Kas}, and 
$\mathrm{R}, \mathrm{R}^\vee, \mathrm{R}^{\!\vee\vee}$
denote the classical part of the 
{\em combinatorial $R$}'s defined in eqs.(2.1), (2.2) and (2.4) in \cite{KOY}, respectively.
They are nontrivial bijections $B_m \times B_l \rightarrow B_l \times B_m$
obeying the Yang-Baxter equations \cite[eq.(2.7)]{KOY}.
The quantities $P_i(\alpha,\beta), Q_i(\alpha, \beta)$ are 
versions of {\em energy functions} 
and known to play an important role \cite{KMN1, NY, KOY}.

As for the $K$ matrix (\ref{ykw}), it has the following behavior at $q=-p^2=0$:
\begin{align}\label{cie}
\lim_{q\rightarrow 0}K'(z)_\alpha^\gamma/
K'(z)_{l{\bf e}_2}^{l{\bf e}_1} = 
\theta(\gamma = \sigma(\alpha))z^{\alpha_1}.
\end{align}
The denominator here can be written down explicitly from
(\ref{akn}) and (\ref{ykw}).
The transformation $\alpha \mapsto \gamma=\sigma(\alpha)$ 
viewed as a bijection on $B_l$ essentially reproduces 
the combinatorial $K$ matrix introduced in \cite[eq.(2.8)]{KOY}
to formulate the box-ball system with reflecting end.
Together with the combinatorial $R$'s in the above,
it forms a set-theoretical solution to the reflection equation.
The latter is known to admit a further generalization to the birational maps
\cite[App.A]{KOY}.
We conclude that the reflection equation (\ref{sin2}), 
after exchange of the two components, achieves a {\em $q$-melting} of 
the combinatorial reflection equation 
\cite[eq.(2.13)]{KOY}.

\begin{example}
Let $n=5$.
We denote $v_{(2,1,0,2,0)} \in V_{5,z}$ by one-row semistandard tableau
$11244$ and similarly $v^\vee_{(0,1,0,3,1)} \in V^\vee_{5,z}$ by 
$\bar{2}\bar{4}\bar{4}\bar{4}\bar{5}$, etc.
With a proper normalization at $q=0$, the action of the two sides of 
$(\ref{sin2})_{x=y=1}$ 
on a base vector $12235 \otimes 124 \in V_{5,1}\otimes V_{3,1}$
proceed, according to (\ref{msa1})--(\ref{cie}),  as follows:
\begin{align*}
12235 \otimes 124 &\overset{R}{\longmapsto} 235 \otimes 11224 
\overset{K'_1}{\longmapsto} \bar{1}\bar{2}\bar{4} \otimes 11224 
\overset{R^\vee}{\longmapsto} 11235 \otimes \bar{1}\bar{3}\bar{5}
\overset{K'_1}{\longmapsto} \bar{1}\bar{2}\bar{4}\bar{5}\bar{5} \otimes \bar{1}\bar{3}\bar{5},
\\
12235 \otimes 124
&\overset{K'_1}{\longmapsto} \bar{1}\bar{1}\bar{2}\bar{4}\bar{5}\otimes 124
\overset{R^\vee}{\longmapsto} 135 \otimes \bar{1}\bar{1}\bar{3}\bar{5}\bar{5}
\overset{K'_1}{\longmapsto} \bar{2}\bar{4}\bar{5}
\otimes \bar{1}\bar{1}\bar{3}\bar{5}\bar{5}
\overset{R^{\vee\vee}}{\longmapsto}
\bar{1}\bar{2}\bar{4}\bar{5}\bar{5} \otimes \bar{1}\bar{3}\bar{5}.
\end{align*}
The agreement of the output 
is an example of the set-theoretical reflection equation \cite{KOY}.
\end{example}

\section{Summary and outlook}\label{sec:d}

In Theorem \ref{th:tgm} 
we have characterized a $K$ matrix as the intertwiner of the 
coideal subalgebra $\mathcal{B}_q$ of $U_q(A^{(1)}_{n-1})$ 
generated by (\ref{slk}). 
By construction, it satisfies the reflection equation (\ref{sin}).
In Theorem \ref{th:ykn} we have constructed it 
in a matrix product form in terms of
terminating $q$-hypergeometric series of $q$-boson generators.

At $q=0$, the $K$ matrix here 
reproduces one of the set-theoretical $K$ matrices called 
``Rotateleft" in \cite[eq.(2.10)]{KOY}.
When $n$ is even, 
there are further solutions known as ``Switch$_{1n}$"  
and ``Switch$_{12}$" \cite[eqs.(2.11), (2.12)]{KOY}
which also admit decent generalizations into geometric versions \cite[app.A]{KOY}.
To incorporate them into the framework of this Letter, possibly with 
some other coideal subalgebra, is a natural problem to be addressed.
Another important theme is to explore the 3D aspects of the matrix product 
(Theorem \ref{th:ykn}) from the viewpoint of \cite{KP}.
It amounts to embedding the relations among the operators 
$G^j_i$ (\ref{aim}) into some sort of  
{\em quantized} reflection equation.
We hope to report on these issues elsewhere.

\section*{Acknowledgments}
The authors thank Vladimir Mangazeev, Zengo Tsuboi and Bart Vlaar for comments.
A.K. is supported by 
Grants-in-Aid for Scientific Research 
No.~18H01141 from JSPS.
M.O. is supported by Grants-in-Aid for Scientific Research No.~15K13429
and No.~16H03922 from JSPS.

\end{document}